\documentclass[superscriptaddress,amsfonts,amssymb,amsmath,twocolumn]{revtex4}

\usepackage{epsfig}
\usepackage{graphics}  
\usepackage{amsmath}
\usepackage{color}
\date{\today}

\begin{document}

%%%%%%%%%%%%%%%%%%%%%%%%%%%%%%%%%%%%%%%%%%%%%%%%%%%%%%%%%%%%%%%%%%%%%%%%%%%%%%%%%%%%%%%%%%%%%%%%%%%%%%%%%%%%%%%%%%%%%%%%%%%%%%%%%%%%%%%%%%%%%%%%%%%%%%%%
%                       Title   

\title{Observer dependence of entanglement in ``nonrelativistic'' quantum mechanics}
    
%%%%%%%%%%%%%%%%%%%%%%%%%%%%%%%%%%%%%%%%%%%%%%%%%%%%%%%%%%%%%%%%%%%%%%%%%%%%%%%%%%%%%%%%%%%%%%%%%%%%%%%%%%%%%%%%%%%%%%%%%%%%%%%%%%%%%%%%%%%%%%%%%%%%%%%%
%                AUTHORS AND ADDRESSES

\author{Matheus H. Zambianco}\email{matheus.hrabowec@unesp.br}    
\affiliation{Instituto de F\'\i sica Te\'orica, 
Universidade Estadual Paulista, Rua Dr.\ Bento Teobaldo Ferraz, 271, 01140-070, S\~ao Paulo, S\~ao Paulo, Brazil}

\author{Andr\'e G.\ S.\ Landulfo}\email{andre.landulfo@ufabc.edu.br}
\affiliation{Centro de Ci\^encias Naturais e Humanas,
Universidade Federal do ABC, Avenida dos Estados, 5001, 09210-580, Santo Andr\'e, S\~ao Paulo, Brazil}

\author{George E.\ A.\ Matsas}\email{george.matsas@unesp.br}
\affiliation{Instituto de F\'\i sica Te\'orica, 
Universidade Estadual Paulista, Rua Dr.\ Bento Teobaldo Ferraz, 271, 01140-070, S\~ao Paulo, S\~ao Paulo, Brazil}

    
%%%%%%%%%%%%%%%%%%%%%%%%%%%%%%%%%%%%%%%%%%%%%%%%%%%%%%%%%%%
%%%%%%%%%%%%%%%%%%%%%%%%%%%%%%%%%%%%%%%%%%%%%%%%%%%%%%%%%%% 
%             ABSTRACT

\begin{abstract}
It was recently shown that, in general, the von Neumann spin entropy of fermionic particles is not invariant under Lorentz boosts. 
We show that an analogous result can be recovered (at the lowest order of $v^2 /c^2$)  using plain {\em nonrelativistic} 
quantum mechanics provided one uses that {\em energy weighs}: $E=m c^2$. This should (i)~help to moderate the 
skepticism on the observer-dependence of the spin entropy of fermionic particles, (ii) emphasize the ``soft'' relativistic 
nature of this result, and (iii)~show that this is a particular case of a more general class of systems, since our calculation 
only assumes a nonrelativistic particle endowed with an internal degree of freedom.           
\end{abstract}

%%%%%%%%%%%%%%%%%%%%%%%%%%%%%%%%%%%%%%%%%%%%%%%%%%%%%%%%%%
%%%%%%%%%%%%%%%%%%%%%%%%%%%%%%%%%%%%%%%%%%%%%%%%%%%%%%%%%%         

\maketitle

%%%%%%%%%%%%%%%%%%%%%%%%%%%%%%%%%%%%%%%%%%%%%%%%%%%%%%%%%%
%%%%%%%%%%%%%%%%%%%%%%%%%%%%%%%%%%%%%%%%%%%%%%%%%%%%%%%%%%
%        INTRODUCTION
\section{Introduction}
\label{sec:Introduction} 

Recently, Peres et al have shown that the von Neumann spin entropy of  a massive fermion is observer 
dependent~\cite{PST02}.  They considered a pure fermionic wave-packet state separable in spin and 
momentum in some inertial frame $I$ having, thus, a vanishing spin entropy $S[\rho^{s}_I]=0$ in such a 
frame, where $\rho^{s}_I$ is  the reduced density matrix  coming from tracing out the momenta degrees 
of freedom. It happens, however, that in a distinct inertial frame $I'$, related to $I$ through a Lorentz boost, 
a Wigner rotation leads spin and momenta to be nonseparable, in general, resulting in a nonzero spin entropy: 
$S[\rho^{s}_{I'}]\neq 0$. 

Soon after, however, some claims were raised against such a conclusion~\cite{C05} and doubts on the standard 
procedure of tracing out the momentum degrees of freedom from the full state to obtain the reduced spin density 
matrix were posted~\cite{RC19}. In particular, Ref.~\cite{SV12} claims that it would not be possible to measure 
the particle spin independently of its momentum in a relativistic setting. Doubts on Peres et al's result based on 
the principles of relativity can be also found~\cite{C10}.  Ref.~\cite{H-Yetal13}, e.g., criticizes Ref.~\cite{PST02}
because {\em the theory of relativity would require a physical quantity to be Lorentz-invariant}. This is incorrect, since 
relativistic observables are in general observer dependent. (The relativity  principle only demands that distinct inertial 
observers measure the same value for any given observable  provided  {\em the experiments are carried on with  states
equally prepared in the corresponding proper frames}; clearly, if identical separable states are prepared in $I$ and $I'$, 
Peres et al would have obtained  $S[\rho^{s}_{I}] = S[\rho^{s}_{I'}] = 0$.) 

As it can be seen, the issue is not consensual yet  and efforts trying to reconcile all pieces of information are in 
course~\cite{TA13}-\cite{C-RN-A15}. This is very much in order, since  the objections to Ref.~\cite{PST02} collide 
with papers according to which Peres et al's  conclusions should impact on our knowledge about the spin correlation of 
entangled fermions as measured by moving detectors~\cite{GA02}-\cite{LM09}. (See also Ref.~\cite{LMT10} where 
fermions are replaced by photons, which might have some practical application in the future.) This paper follows this trend by 
looking for a simple ``nonrelativistic'' analogous  system, where similar conclusions as the ones obtained in 
Ref.~\cite{PST02} can be extracted.  We consider a particle endowed with an internal-energy degree 
of freedom in an  one-dimensional box with boundary conditions that allow the existence of momentum eigenstates.  
The momentum eigenstates will be denoted by 
$|p\rangle$, as usual, while the internal-energy eigenstates will be given by $|E_1 \rangle$  and $|E_0\rangle$,   
$E_1>E_0.$  Such an internal degree of freedom can be realized, for instance, by coupling 
a nonrelativistic spin to a constant magnetic field ${\bf B}$. The nonexcited, 
$|E_0 \rangle\equiv |\leftarrow \,\rangle$, and excited, $|E_1\rangle \equiv |\rightarrow \,\rangle$, 
energy eigenstates can be associated with the spin states being aligned and counteraligned with  ${\bf B}$, respectively, 
which is  conveniently set up to point out along the boost direction.  We recall that the magnetic field component 
along the boost direction is invariant: ${\bf B}_{||}={\bf B}'_{||}$. 

In analogy to Ref.~\cite{PST02}, we prepare the state to be separable in some inertial frame~$I$ as
\begin{equation}
    |\psi_I\rangle=|\uparrow\, \rangle \otimes |p\rangle,
    \label{stateI}
\end{equation}
where we have defined the orthogonal basis
$$
|\uparrow\,\rangle \equiv\frac{1}{\sqrt{2}}( |\leftarrow\, \rangle + |\rightarrow\rangle),\quad 
|\downarrow\,\rangle \equiv \frac{1}{\sqrt{2}}(|\leftarrow\, \rangle - |\rightarrow\rangle).
$$
Obviously, $ |\psi_I\rangle$  has vanishing spin entropy $S[\rho^{s}_I]=0$, where 
$\rho^{s}_I \equiv  {\rm Tr}_p |\psi_I\rangle \langle\psi_I |$. 

We wonder what is the entropy $S[\rho^{s}_{I'}]$ as defined in an inertial frame $I'$, related to 
$I$  by a Galilean boost. We will show that $S[\rho^{s}_{I'}]\neq 0$ provided one uses that  
$E=m c^2$. We also verify that by taking the full nonrelativistic limit at the end, i.e. $c \to \infty $, we recover the commonsensical 
conclusion that the momentum and internal-degree-of-freedom entanglement is invariant under Galilean 
transformations~\cite{PC03}-\cite{H07}:  $S[\rho^{s}_{I'}]_{c\to \infty} = 0$.  This should help to moderate the skepticism on the 
observer-dependence of the spin entropy for fermionic particles, emphasize the ``soft'' relativistic nature of this
result, and point out that this is a particular case of a more general class of systems. 

The paper is organized as follows. In Sec.~\ref{sec:Lorentz}, we review Peres et al's main results~\cite{PST02}. 
In Sec.~\ref{sec:Galileo} we show how analogous results can be obtained in a nonrelativistic setting provided one recalls
that energy weighs. Our closing remarks appear in Sec.~\ref{sec:Conclusions}. We assume $\hbar=1$ but keep $c$ 
in our formulas for the sake of clarity.

%%%%%%%%%%%%%%%%%%%%%%%%%%%%%%%%%%%%%%%%%%%%%%%%%%%%
%%%%%%%%%%%%%%%%%%%%%%%%%%%%%%%%%%%%%%%%%%%%%%%%%%%%
% Peres
\section{Spin entropy for a fermionic particle under Lorentz boosts}
\label{sec:Lorentz} 

Here, we review the main results of Ref.~\cite{PST02} for the sake of further comparison. Let us start considering 
a general spin-$1/2$ fermionic particle described by a spinor in the momentum representation as
\begin{align}
   \psi ({\bf p} ) \equiv
        \begin{bmatrix}
            a_1 ({\bf p})  \\ 
            a_2 ({\bf p}) 
        \end{bmatrix}.
        \label{psi(p)}
\end{align}
Tracing out the momentum degrees of freedom, the spin entropy of the reduced density matrix can be written as 
\begin{equation}
S[\rho^{s}] =-\frac{1-|{\bf n} | }{2} \log\left(\frac{1-|{\bf n} | }{2} \right) -\frac{1+ |{\bf n} |}{2} \log\left(\frac{1+|{\bf n} |}{2}\right),
\label{SPeres}
\end{equation} 
where ${\bf n} =(n^x, n^y, n^z)$ is the Bloch vector with 
\begin{equation}
n^z=\int d^3{\bf p} (|a_1|^2-|a_2|^2),
\label{n_z}
\end{equation}
\begin{equation}
n^x -i n^y=2\int d^3{\bf p} \;a_1 a_2^* .
\label{n_x,n_y}
\end{equation}

Now, let us consider a particular case where the particle state 
\begin{align}
   \psi_I ({\bf p} ) \equiv
        \begin{bmatrix}
             \left({1}/{ \pi w^2} \right)^{3/4} \exp\left(-{|{\bf p}|^2}/{2 w^2} \right)\\ 
             \\
            0
        \end{bmatrix}
        \label{psi_I(p)}
\end{align}
 is  a  Gaussian wavepacket (in momentum space) with width $w={\rm const}$   and spin $+1/2$ along 
 the $z$ axis with respect to a congruence of observers lying at rest in the inertial frame $I$. In this case,  
 ${\bf n}=(0,0,1)$ and one obtains $S[\rho^{s}_{I}]=0$ from Eq.~(\ref{SPeres}), as expected. 

Next, let us wonder how $\psi ({\bf p} )$  looks like for a different congruence of observers lying at rest 
in the inertial frame $I'$, which is Lorentz boosted along the $x$ direction with velocity $v$. After 
performing a Wigner rotation, Peres et al obtain 
 \begin{equation}
   \psi_{I'} ({\bf q} ) 
   \equiv
        \begin{bmatrix}
             {a_1}'({\bf q})\\ 
             {a_2}'({\bf q})          
        \end{bmatrix}
  \equiv
        K a_1(\mathbf{p}) 
        \begin{bmatrix}
             b_1({\bf p})\\ 
             b_2({\bf p})            
        \end{bmatrix},
        \label{psi_I'(p)}
\end{equation}
where the 3-momenta ${\bf p}$ and ${\bf q}$ are related to each other through 
$p^{\mu}=(\Lambda^{-1} q)^{\mu}$ with $\Lambda$ being the boost matrix. Here,
\begin{eqnarray}
 a_1( \mathbf{p}) &=&  \left({1}/{ \pi w^2} \right)^{3/4} \exp\left(-{|{\bf p}|^2}/{2 w^2} \right),
   \label{a1}
\nonumber   \\
 b_1( \mathbf{p}) &=& \cosh{\left(\frac{\alpha}{2}\right)}(p^0+mc) - \sinh{\left( \frac{\alpha}{2} \right)}(p^x+ip^y)
\label{b1}
\nonumber \\
b_2(\mathbf{p}) &=&  -\sinh{\left( \frac{\alpha}{2} \right)}p^z,
\nonumber
\label{b2}
\end{eqnarray}
with $\tanh \alpha \equiv \beta\equiv v/c$ and 
\begin{equation}
   K = \left[\frac{p^0}{q^0 (p^0+mc)(q^0+mc)} \right]^{1/2}.
   \label{K}
\end{equation}
Then, by using Eq.~(\ref{psi_I'(p)}) in  Eqs.~(\ref{n_z})-(\ref{n_x,n_y}) (with $a_i \to {a_i}' $, $i=1,2$) and recalling that  
\begin{equation}
d^3\mathbf{p}/p^0= d^3\mathbf{q}/q^0, 
\end{equation}
we get the transformed Bloch vector:
$\mathbf{n}'=({n^x}', {n^y}', {n^z}')$, where 
\begin{equation}
{n^x}'={n^y}'=0,\quad  {n^z}'=\int{d^3 \mathbf{r} \  \frac{\exp (-{ |{\bf r}|^2}/{{\widetilde w}^2} )}{{\widetilde w}^3 \pi^{{3}/{2}}}G( \mathbf{r})},
  \label{NZ}
\end{equation}
with ${\widetilde w} \equiv w/mc$, and we have performed the replacement ${\bf q} \to {\bf r} \equiv {\bf q}/mc$. Moreover, 
\begin{equation}
G(\mathbf{r}) =
\frac{ (\gamma + 1 -\gamma \beta x)(1 +\sqrt{1+ |{\bf r}|^2})+\gamma(x^2 + y^2) + z^2}
{(1+\sqrt{1+|{\bf r}|^2})[1+\gamma(\sqrt{1+|{\bf r}|^2}-\beta x)]},
\label{G}
\end{equation}
where 
\begin{equation}
x\equiv |{\bf r}| \sin \theta \cos \phi, \,  y\equiv |{\bf r}| \sin \theta \sin \phi, \, z\equiv|{\bf r}| \cos \theta,
\end{equation} 
with
$0\leq \theta \leq \pi$, $0\leq \phi < 2\pi$ and $\gamma = (1-\beta^2)^{-1/2}$.

In order to exhibit more clearly the physical content of this result, it is convenient to  expand  $G(\mathbf{r})$ 
around $\mathbf{r}=0$ before we evaluate the integral in Eq.~(\ref{NZ}). The output comes out automatically as 
a series for  ${\widetilde w}$:
\begin{eqnarray}
 {n^z}'
 & = &
 1
 - \left(\frac{\gamma -1}{\gamma +1} \right)\frac{{\widetilde w}^2}{4} 
 +\frac{(11 \gamma^3 +9\gamma^2 -11\gamma -9)}{(1+\gamma)^3}\frac{{\widetilde w}^4}{32}
 \nonumber \\
 & + & {\cal O} (\widetilde{w}^6) 
 \label{array1} \\
& = &
 1- \left( \frac{\widetilde{w}^2}{16}  - \frac{5 \widetilde{w}^4}{64} + 
{\cal O}(\widetilde{w}^6)\right)\beta^2 + {\cal O}(\beta^4),
\label{nz(v)}
\end{eqnarray}
where the second equality comes from an extra expansion for $\beta \ll 1$, which will be useful later.

For sharp momentum states, i.e., $\widetilde{w} \ll 1$, the first terms of  Eq.~(\ref{array1}) 
approximate ${n^z}'$ very well. Then, Peres et al write the spin entropy at leading order in $\widetilde{w}^2$ as
 \begin{equation}
 S[\rho^{s}_{I'}] \approx t(1-\log t),
 \label{SP}
 \end{equation}
 with
 \begin{equation}
  t=\frac{\widetilde w^2}{8}\left(\frac{\gamma -1}{\gamma +1} \right).
  \label{t}
\end{equation}  
Clearly, $ S[\rho^{s}_{I'}] \neq 0$ provided $v\neq 0$ and the wave packet is not arbitrarily sharp: $\widetilde w \neq0$.

%%%%%%%%%%%%%%%%%%%%%%%%%%%%%%%%%%%%%%%%%%%%%%%%%%%%
%%%%%%%%%%%%%%%%%%%%%%%%%%%%%%%%%%%%%%%%%%%%%%%%%%%%
% Galileo
\section{Spin entropy for a nonrelativistic particle under Galilean boosts}
\label{sec:Galileo} 

 Let us consider, now, a free particle with rest mass $M$ endowed with an internal-energy degree of freedom and 
 constrained to  move in an one-dimensional box with size $L$. The Hilbert space of our system is 
 ${\cal H}= {\cal H}_p \otimes \cal H_{E}$, where the momentum and 
 internal-energy Hamiltonian operators act as follows:  $\hat{p}:{\cal H}_{p} \to {\cal H}_{p}$ and 
 ${\hat{H}}_{\cal E}: \cal H_{E} \to \cal H_{E}$, respectively~\cite{Note1}.  The total Hamiltonian  is simply
 \begin{equation}
   \hat{H}=\frac{\hat{p}^2}{2M} + \hat{H}_{ \cal E}. 
   \label{Hamiltoniana}
 \end{equation}

Let us prepare our state to be separable in some inertial frame $I$ at some instant $t=0$ as
\begin{eqnarray}
|\xi_I  \rangle  & = &\frac{1}{\sqrt{2}}( |E_0 \rangle + |E_1\rangle) \otimes | p_n \rangle 
\nonumber \\
& = &\frac{1}{\sqrt{2}}( |E_0, p_n  \rangle + |E_1,p_n \rangle ),
\label{xiI}
\end{eqnarray}
where  the nonexcited $|E_0 \rangle$,  excited $|E_1\rangle$, and momentum $|p_n \rangle$ eigenstates   satisfy 
$$
\hat{H}_{\cal E} |E_0 \rangle=E_0 | E_0 \rangle, \; \hat{H}_{\cal E} |E_1 \rangle= E_1 | E_1\rangle,
$$ 
and 
$$
\hat{p}|p_n \rangle= p_n |p_n \rangle,
$$
respectively. Here, $p_n=2 \pi n/L$ for $n \in \mathbb{Z}$, once 
we have assumed that the particle wave function obeys periodic boundary conditions. 
Superposition states of internal energy levels are routinely produced in laboratory (see, e.g., 
Ref.~\cite{CHRW10} for a recent application concerning the measurement of time dilation).

Clearly, the reduced spin matrix 
$$
\rho^{s}_{I}= {\rm Tr}_p |\xi_I  \rangle \langle\xi_I  |
$$ 
obtained after tracing out the momenta degrees of freedom still represents a pure quantum state, leading to  a vanishing von Neumann entropy: 
$$
S[\rho^{s}_{I}]=0.
$$ 
The label ``$s$'' appears because we have associated the nonexcited and excited states to nonrelativistic spin states (as discussed in Sec.~\ref{sec:Introduction}): 
$$
|E_0 \rangle \mapsto |\leftarrow\, \rangle,\;  |E_1\rangle \mapsto |\rightarrow\, \rangle,
$$
in order to keep the nonrelativistic analysis as close as possible to the relativistic one. 

Next, we consider the same quantum system as seen in an inertial frame $I'$ related with $I$ through a 
Galilean boost along the $x$ direction with velocity $v$. For this purpose, we recall that the wave function 
of the system at $t=0$ should be transformed by the boost operator~\cite{KT03}-\cite{Note2}
\begin{equation}
\hat{G}(v,M)= \exp (iMv \hat{x}),
\label{generalboost}
\end{equation}
where $M$ is the corresponding particle rest mass and $\hat{x}$ is the position operator.  Thus, the boosted  state will be written as
\begin{eqnarray}
|\xi_{I'}  \rangle 
& = & \frac{1}{\sqrt{2}}|E_0\rangle \otimes  \hat{G}  (v,M_0) |p_n  \rangle 
\nonumber \\
& + & \frac{1}{\sqrt{2}}|E_1\rangle \otimes  \hat{G}  (v,M_1) |p_n  \rangle
\nonumber \\
& = & \frac{1}{\sqrt{2}}  |E_0\rangle \otimes |p_n +M_0v \rangle 
\nonumber \\
& + & \frac{1}{\sqrt{2}}  |E_1 \rangle \otimes  |p_n +M_1v \rangle
\label{xiNovo}
\end{eqnarray}
where $M_{j}= \displaystyle m + {E_j}/{c^2}$ for $j=0,1$. Here, we have used that mass is what scales measure; and scales measure 
the total energy of the system in its rest frame. Hence, the internal energy contribution $E_j/c^2$  must be added to the bare mass $m$ 
(i.e. the mass that the particle would have without any internal energy) to give $M_j$. 

As a side comment, we note that Bargamann's celebrated result that the Galilean group imposes a superselection rule  in nonrelativistic 
quantum mechanics precluding the superposition of distinct mass eigenstates~\cite{B54} does not apply here for two reasons. From the 
mathematical side,  Eq.~(\ref{xiNovo})  solely relies  on the one-dimensional subgroup of Galilean boosts (rather than on the whole 
Galilean group), which does not lead to any such a superselection rule. Bargmann's superselection rule comes from considering the composition 
of a sequence of boosts {\em and translations}. Furthermore, by using $E=mc^2$, we make it explicit that our system enherits elements of 
relativity, which drives it beyond the scope of Bargmann's theorem (for a more detailed discussion on it see Ref.~\cite{G01}). From 
the physical side, it is consensual that the superposition of mass eigenstates is realized by nature, being the basis, e.g., of neutrino 
oscillation experiments. 

It is also interesting to note that the same result~(\ref{xiNovo}) can be obtained by writing 
\begin{equation}
|\xi_{I'} \rangle = \hat{{\cal G}}(v)|\xi_{I} \rangle,
\label{xiI'}
\end{equation}
with the unitary operator
\begin{equation*}
\hat{{\cal G}}(v) = \exp (i \hat{M} v \hat{x}),
\label{Gcal}
\end{equation*}
where $M$ in Eq.~(\ref{generalboost}) is promoted to the Hermitian operator 
\begin{equation}
\hat{M}=m\hat{\cal I_{E}} + {\hat{H}_{\cal E}}/{c^2}.
\label{M}
\end{equation}
It is worthwhile to note that the set of unitary operators $\hat{{\cal G}}(v)$ gives rise to a faithful representation of the one-dimensional 
Galilean boost subgroup:
$$
g_{v_1} \circ g_{v_2} = g_{v_1 +v_2}   \implies \hat{{\cal G}}(v_1) \hat{{\cal G}}(v_2) = \hat{{\cal G}}(v_1+ v_2)
$$
with $\hat{{\cal G}}(0)  = \hat{\cal I}$. The fact that Eq.~(\ref{xiI'}) coincides with Eq.~(\ref{xiNovo}) can be straightforwardly checked 
out by noting that 
\begin{equation}
\hat{{\cal G}}(v)|E_{j}, p \rangle= |E_{j} \rangle \otimes \hat{G}(v,M_j)|p \rangle.
\label{GcalEX}
\end{equation}
As a consequence,  $M$ in Eq.~(\ref{Hamiltoniana}) should be also promoted  to $\hat M$ for the sake of consistency. 
The present analysis does not involve dynamics and, thus, our results are insensitive to such a promotion. Situations which do involve 
dynamics are much subtler. See, e.g., Ref.~\cite{PZCB15} for a related case, where the system is time evolved in a gravitational 
field and the thrilling debate which was sparked from it~\cite{PCK16}.

Now, we compute the density matrix for the system as a whole and then trace out the momentum degrees of freedom:
 \begin{eqnarray}
  \rho_{I'}^{s}(v) 
  &\equiv&  {\rm Tr}_p |\xi_{I'} \rangle \langle\xi_{I'}  |
 \nonumber \\
 &=&
 \frac{1}{2}(|E_0 \rangle\ \langle E_0|+|E_1 \rangle\ \langle E_1|)
 \nonumber \\
& + & 
 \frac{1}{2}( f(v)|E_0 \rangle\langle E_1|  + f(v)^* |E_1 \rangle\ \langle E_0|),
 \label{RangoLango}
 \end{eqnarray}
 where
 \begin{eqnarray}
 f(v)
 & = & {\rm Tr}_p (|p_n +M_0v \rangle \langle p_n +M_1 v |)
 \nonumber \\
 & = & \exp ( i v \epsilon   L /2 c^2)  \frac{\sin (v L \epsilon/2 c^2)}{(v L \epsilon/2 c^2)}
  \label{f(v)}
 \end{eqnarray}
with $\epsilon \equiv (E_1 - E_0)$. We note that $\lim \limits_{v \to 0} f(v)=1$ implies
$\lim \limits_{v \to 0} \rho_{I'}^{s}(v)=\rho_{I}^{s}$, as expected.
Finally, the von Neumann entropy  reads
\begin{equation}
 S[\rho^{s}_{I'}]
= -  \frac{1-|f|}{2} \log{\left( \frac{1-|f|}{2}\right)} - \frac{1+ |f|}{2} \log{\left(\frac{1+ |f|}{2} \right)},
 \label{array2}
\end{equation}
where
\begin{eqnarray}
 |f| &=&  \frac{\sin (v L \epsilon/2 c^2)}{(v L \epsilon/2 c^2)},
  \label{g(v)}   \\
&=& 1-\left(\frac{\epsilon^2 L^2}{24 c^2} \right) \beta^2 +{\cal O}(\beta^4).
\label{gvTaylor}
\end{eqnarray}
In the last step, we have written $\beta = v/c$. 

It is clear from Eq.~(\ref{g(v)}) that $S[\rho^{s}_{I'}] \neq 0$ provided $\epsilon \neq 0$. Most interestingly,  
in this nonrelativistic limit, there is no doubt that one can measure the internal degree of freedom (e.g., spin) 
independently of its momentum. Hence, Eqs.~(\ref{array2})-(\ref{gvTaylor}) show that the noninvariance of the 
spin entropy in different frames is not a technical artifact but a physical fact which may influence experimental outputs
when the observer who prepares the state and the one who measures it move with respect to each other.   

Let us finish by comparing the relativistic and nonrelativistic spin entropy results. To this end, we note that both  
expressions~(\ref{SPeres}) and~(\ref{array2}) are formally the same and thus it is enough to
compare our $|f|$ in Eq.~(\ref{gvTaylor}) with ${n^z}'$ in Eq.~(\ref{nz(v)}). We see that they are comparable 
at the lowest order of $\beta^2$ provided one makes the identification 
\begin{equation}
\frac{\epsilon^2 L^2}{24 c^2} \leftrightarrow  \frac{\widetilde{w}^2}{16}  - 
\frac{5 \widetilde{w}^4}{64} + {\cal O}(\widetilde{w}^6).
\label{identification1}
\end{equation}
The identification above can be recast in a more suitable form by introducing the electron Compton wavelength, 
$\lambda= 2\pi/(m c)$ ($\hbar=1$):
\begin{equation}
\frac{(\epsilon/c)^2 L^2}{24} \leftrightarrow \frac{ w^2 \lambda^2}{64 \pi^2}, 
\label{identification2}
\end{equation}
where we have kept only the leading term in $w$ in the right-hand side of Eq.~(\ref{identification2}), since the result 
in the left-hand side, obtained from our nonrelativistic calculation, comes from  assuming momentum eigenstates. 
Apart from numerical multiplicative factors, $w$ and $\lambda$ are seen to play the role of $\epsilon/c$ and $L$, respectively.

Despite the similarity between Peres et al and our analyses, they should be seen as being 
complementary to each other in the sense that while they consider {\em free} electrons with momentum width scaled by  $w$, 
we consider a {\em confined} particle in a box with length~$L$ and well defined momentum. 

Finally, we note that the conclusion reached in Refs.~\cite{PC03}-\cite{H07} that the entropy of the reduced density matrix, 
 in the nonrelativistic quantum realm, is invariant under Galileo boosts can be recovered here by simply taking 
 $c \to \infty$ in Eqs.~(\ref{array2})-(\ref{g(v)}): $\lim \limits_{c\to \infty} S[\rho^{s}_{I'}] =S[\rho^{s}_{I}]= 0$;
 no interesting result is obtained unless one recalls that energy weighs.

%%%%%%%%%%%%%%%%%%%%%%%%%%%%%%%%%%%%%%%%%%%%%%%%%%%%
%%%%%%%%%%%%%%%%%%%%%%%%%%%%%%%%%%%%%%%%%%%%%%%%%%%%
% Conclusions

\section{Conclusions}\label{sec:Conclusions} 

We have have considered a nonrelativistic particle endowed with an internal degree of freedom. 
Such a degree of freedom plays the role of the spin of an electron  described by a relativistic fermionic field. 
We prepare our nonrelativistic state to be separable (in spin and momentum) in  some inertial frame~$I$. 
We have shown that it will be nonseparable, in general, in some other inertial frame $I'$ related to $I$  by a 
Galilean boost {\em if} we recall that energy weighs. The spin entropy obtained can be compared with the one 
given in Ref.~\cite{PST02} in the nonrelativistic regime $\beta \ll 1$.  We hope that our paper moderates the 
skepticism concerning the observer dependence of the spin entropy for fermionic particles, since it does
not involve any Wigner rotation which seems to be the core of the dispute. 

%%%%%%%%%%%%%%%%%%%%%%%%%%%%%%%%%%%%%%%%%%%%%%%%%%%%
%%%%%%%%%%%%%%%%%%%%%%%%%%%%%%%%%%%%%%%%%%%%%%%%%%%%

\acknowledgments

 M.~Z.\ and  A.~L.~were fully and partially supported by S\~ao Paulo Research Foundation under 
 Grants  2018/24810-5 and 2017/15084-6, respectively. G.~M.\ was partially supported by Conselho 
 Nacional de Desenvolvimento Cient\'\i fico e Tecnol\'ogico under grant 301544/2018-2.

%%%%%%%%%%%%%%%%%%%%%%%%%%%%%%%%%%%%%%%%%%%%%%%%%%%%%
%%%%%%%%%%%%%%%%%%%%%%%%%%%%%%%%%%%%%%%%%%%%%%%%%%%%%

\end{document}